\documentclass[aps,pre,reprint,groupedaddress,nofootinbib]{revtex4-1}

\usepackage{graphicx}
\usepackage{upgreek}
\usepackage{color}
\begin{document}

%\preprint{}

\title{Bifurcation induced by the aspect ratio in a turbulent Von-K\'arm\'an swirling flow}


\author{Olivier Liot}
 \altaffiliation[]{Univ Lyon, ENS de Lyon, Univ Claude Bernard, CNRS, Laboratoire de Physique, F-69342 Lyon Cedex 7, France}
 \altaffiliation{Presently at LAAS-CNRS, 7 avenue du Colonel Roche, BP54200 31031 Toulouse Cedex 4, France}
\author{Javier Burguete}%
\email[]{javier@unav.es}
\affiliation{Departamento de F\'isica y Matem\'atica Aplicada, Universidad de Navarra, c/ Irunlarrea, Pamplona, Spain}
%\thanks{}


\date{\today}

\begin{abstract}

We evaluate the effect of two experimental parameters on the slow dynamics of a Von-K\'arm\'an swirling flow driven by two propellers in a closed cylinder. 
The first parameter is the inertia momentum of the propellers, and the second parameter is the aspect ratio, i.e. the distance between the propellers $H$ divided by the diameter $D$.
% This paper presents how the distance $H$ between the propellers changes the characteristics of a Von-K\'arm\'an swirling flow. 
We use a cell with a fixed diameter $D$ but where the distance between the propellers can be turned continuously and where the inertia from the propellers can also be changed using different gears. No change on the dynamics is observed when the momentum of inertia is modified. Some dramatic changes of the shear layer position are observed modifying the aspect ratio $\Gamma=H/D$. A bifurcation of the shear layer position appears. Whereas for low $\Gamma$ the shear layer position has a smooth evolution when turning the asymmetry between the rotation frequency of the propellers, for high $\Gamma$ the transition becomes abrupt and a symmetry breaking appears. Secondly we observe that the spontaneous reversals already observed in this experiment for $\Gamma=1$ [de la Torre \& Burguete \textit{PRL} \textbf{99}, 054101 (2007)] exist only in a strait window of aspect ratio. We show using an experimental study of the mean flow structure and a numerical approach based on a Langevin equation with coloured noise that the shear layer position seems to be decided by the mean flow structure whereas the reversals are linked to the spatial distribution of the turbulent fluctuations in the cell. 
\end{abstract}

%\keywords{}

\maketitle


\section{Introduction} 

 Turbulent flows can be encountered at very different scales in nature: from geophysical and astrophysical scales (atmospheric turbulence, star convection) to biological scales (heart valves). That is why turbulence is a central problem of present fundamental researches and applied sciences \cite{tennekes1987}. Because of the non-linearity of the equations which rule turbulence, there is no theoretical nor numerical complete description of this phenomenon. Particularly, lots of questions are still open about the appearance and the dynamics of vortices or other coherent structures in fully developed turbulence, or the rise of bifurcations on the mean flow \cite{holmes1998}.

This paper proposes an experimental investigation in a swirling Von-K\'arm\'an flow. Two counter-rotating propellers are used to develop turbulence in a cylindrical cavity filled with water. This model system has been largely studied numerically \cite{nore2003,nore2006,shen2006}, theoretically \cite{vonkarman1921,zandbergen1987} and experimentally \cite{marie2003,ravelet2005,nore2005,delatorre2007}. This kind of flow is particularly used in magnetohydrodynamics (MHD) experiments. It is a good candidate for the dynamo instability which was observed last decade \cite{monchaux2007}. The counter-rotating swirling flow can be the place of multistability, memory effects and long time dynamics \cite{ravelet2004,delatorre2007,burguete2009,machicoane2014}. Particularly de la Torre and Burguete \cite{delatorre2007,burguete2009} observed a symmetry breaking of the mean flow where the shear layer between the two counter-rotating cells of the flow does not remain in the middle of the cavity. Moreover this shear layer can spontaneously jump from one side of the cavity to the other with a long residence time (typically 1000\,s) compared to the turbulent time-scales. But what is/are the problem parameter(s) which fix(es) the position of the shear layer and the spontaneous reversals?

In this configuration the propellers provide the angular momentum and kinetic energy to the fluctuating turbulent flow. These propellers can be controlled using constant torque or angular velocity, and different regimes are obtained for each approach \cite{ravelet2008}. This interaction between the flow and the propellers can be responsible for the different observed dynamics in various experimental setups. Here we would like to test the effect of two parameters related to this interaction: the momentum of inertia of the propellers and the aspect ratio $\Gamma$ (the ratio between the distance between the propellers and the cavity diameter).

For the first parameter, in a different configuration (a thermoconvective experiment) it has been observed that a feedback can be established between the forcing mechanism (with low thermal inertia) and the dynamical behaviour of the experiment when the system is not able to respond to the fluctuating requirements of the flow \cite{ringuet1993}. In the von K\'arm\'an flow, an equivalent behaviour would be obtained for extremely low momentum of inertia of the propellers if the motors cannot follow the fluctuations. 

Concerning the second parameter we systematically study in this paper the impact of the aspect ratio on the symmetry breaking and the spontaneous long time reversals of the shear layer position. If the mean flow structure fixes the shear layer position, spontaneous reversals are observed only in a short window of the aspect ratio. To explain this phenomenon we use a model based on a Langevin equation to link our experimental results to different shapes of the potential and of the spatial distribution of the turbulent fluctuations. 

\section{Experimental setup} 

\begin{figure}[h]
\begin{center}
\includegraphics[width=8cm]{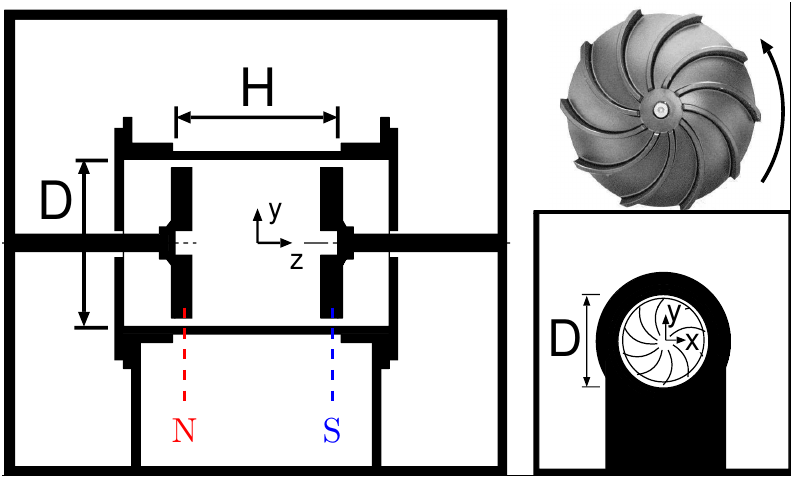}
\caption{Global scheme of the experimental setup. (left) Side view of the cavity with the denomination of the propellers; (top-right) photograph of a propeller with its rotation direction; (bottom-right) side view of the cavity in the propellers axis direction.}
\label{setup}
\end{center}
\end{figure}

 Our Von-K\'arm\'an setup consists in a horizontal cylinder of diameter $D=20$\,cm and of 32\,cm in length filled with water. The distance $H$ between the propellers can be turned continuously. The aspect ratio defined as $\Gamma=H/D$ can be fixed in the range $]0.2,1.2]$ with an accuracy of 0.01. Each propeller has a radius $R_p=8.75$\,cm and has ten curved blades of 2\,cm in tall and 4.85\,cm of curvature radius. The figure \ref{setup} shows a scheme of the experimental setup. The propellers rotate in opposite directions and are powered by two motors of maximum power 1.5\,kW. They are named $N$ and $S$ (see fig. \ref{setup}). The respective rotation frequencies are $f_N$ and $f_S$. We define the rotation asymmetry parameter as $\Delta=(f_N-f_S)/(f_N+f_S)$. The rotation frequency can be adjusted independently for both propellers in a range $[0.5,12.5]$\,Hz with an accuracy of 0.03\,Hz. The high inertia of the propellers leads us to have a very stable rotation frequency with maximal fluctuations of 0.1\%. A higher fluctuation level on the velocity of the propeller will destroy the dynamics observed here \cite{burguete2009,lopez2016}. The Reynolds number is computed as $Re=\pi R_p^2(f_N+f_S)/\nu$ where $\nu$ is the kinematic viscosity of the fluid. We maintain the temperature of the water at $21\pm1^\mathrm{o}$C by immersing the cavity in a tank of 150\,L in volume.

This last is also useful to reduce optical distortions: we make velocity measurements using a 1D Laser Doppler Anemometry (LDA) system placed on a horizontally translating board. Its measurement volume is $1.3\times0.3\times0.3$\,mm$^3$, the temporal resolution is up to 100\,kHz. The flow is seeded with silver coated hollow glass spheres with a diameter of 14\,$\mu$m and a density of 1.65\,g/cm$^3$. Two kinds of measurements are made. (i) Equatorial (orthoradial) velocity measurements at the center of the cell ($y=0,~z=0$ and radial position $y=0.9\,R$) are used to study the shear layer position. (ii) Measurements of 2D velocity fields by the combination of the equatorial and axial velocity measurements in a plane included between the propellers blades extremities and $y\in [0,R]$ are used to estimate the poloidal-toroidal velocity ratio of the mean flow. The volume average of the toroidal and poloidal velocities are respectively defined as: 

\begin{equation}
\left\lbrace
\begin{array}{l}
\overline{V_{tor}}=\frac{1}{\mathcal{V}}\displaystyle{\int_\mathcal{V}}|v_{eq}|d\mathcal{V},\\
~~\\
\overline{V_{pol}}=\frac{1}{\mathcal{V}}\displaystyle{\int_\mathcal{V}}\sqrt{v_{r}^2+v_{z}^2}d\mathcal{V},\\
\end{array}
 \right. 
 \label{eq:V}
\end{equation}

\noindent where $v_{eq}$ is the equatorial velocity, $v_r$ the radial one, $v_z$ the axial one and $\mathcal{V}$ the volume of the cell between the propellers.

\begin{figure}[h!]
\begin{center}
\includegraphics[width=7.5cm]{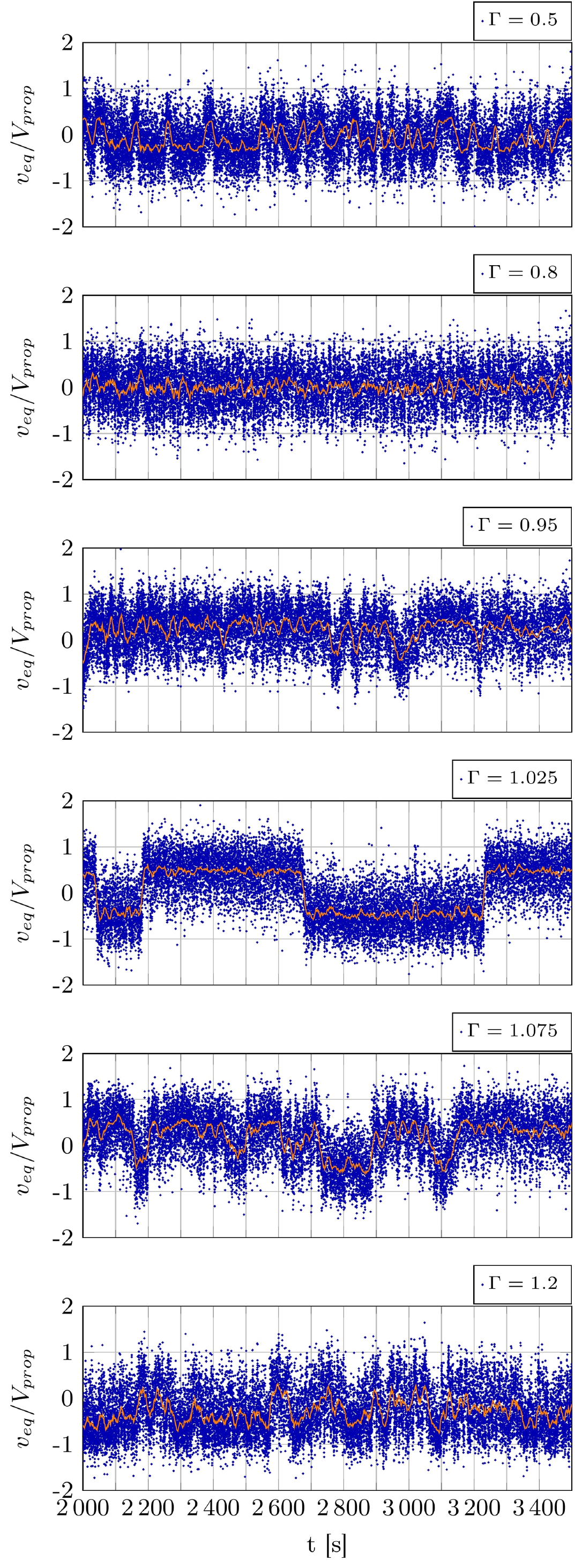}
\caption{Time-series of the equatorial velocity at ($z=0$, $y=0.9\,R$) for six different aspect ratio. $V_{prop}$ represents the velocity of the propellers at the blades outlet. The orange solid lines represent the sliding mean value of the velocity signal. $Re=2.9\times10^5$ and $\Delta \cong 0$.}
\label{timeseries}
\end{center}
\end{figure}

\section{Experimental observations} 

 We first verified that the Reynolds number ($Re$) % and the inertia of the propellers 
does not have any effect on the shear layer position. This was expected as it has already been shown that this configuration does not depend on the Reynolds for $Re>10^5$ \cite{delatorre2007,ravelet2008}.

In previous experiments \cite{delatorre2007,burguete2009} the propellers were connected with a system of gears and belts to the motors, with a total momentum of inertia of $I_{prop}=55.28\times 10^{-3}$\,kg.m$^2$. Here we have used two supplementary configurations that reduce the inertia of each propeller to $I_{prop}=25.76$ and $9.84\times10^{-3}$\,kg.m$^2$. Just for comparison, the momentum of inertia for a solid rotation of the fluid present in each one of the half-cells is $15.71\times10^{-3}$\,kg.m$^2$. For these values of $I_{prop}$ we have not observed any modification on the dynamics. However, we would like to highlight that these values are probably too important and we should test smaller inertias to verify that this parameter has no effect on the dynamics.

The influence of the aspect ratio was evaluated measuring the equatorial velocity at the center of the cell. Figure \ref{timeseries} presents time-series of this velocity over 1500\,s for $\Gamma=0.5,0.8,0.95,1.025,1.075,1.2$. The corresponding Reynolds number reaches $2.9\times10^5$ whereas $\Delta$ is fixed as close to zero as possible ($\Delta\lesssim 10^{-3}$). We observe various behaviours. For $\Gamma=0.5$, whereas the mean velocity is null, some changes of the velocity sign are observed with a typical time scale of some tens of seconds. At $\Gamma=0.8$ the velocity is centered on zero with only turbulent fluctuations. When we increase the aspect ratio ($\Gamma=0.95,1.2$) we observe that the velocity has a privileged sign and presents some large deviations to the other sign. The spontaneous reversals observed by de la Torre \& Burguete \cite{delatorre2007} are also observed at $\Gamma=1.025$. Finally for $\Gamma=1.075$ such reversals are still observable but with a less clear bistability which indicate the behaviour for higher $\Gamma$. To summarize, we have a short $\Gamma$-window where we can observe long-time reversals. Moreover if the mean velocity is very close to 0 for $\Gamma<0.95$, we do not succeed in observing a zero-mean velocity for higher aspect ratios using series of 1500\,s of acquisition time.

%These fluctuations have nothing to do with the long-time flow reversals and could be linked to the dynamics of vortexes travelling in the equatorial direction along the cylinder wall. 

\begin{figure}[h!]
\begin{center}
\includegraphics[width=8cm]{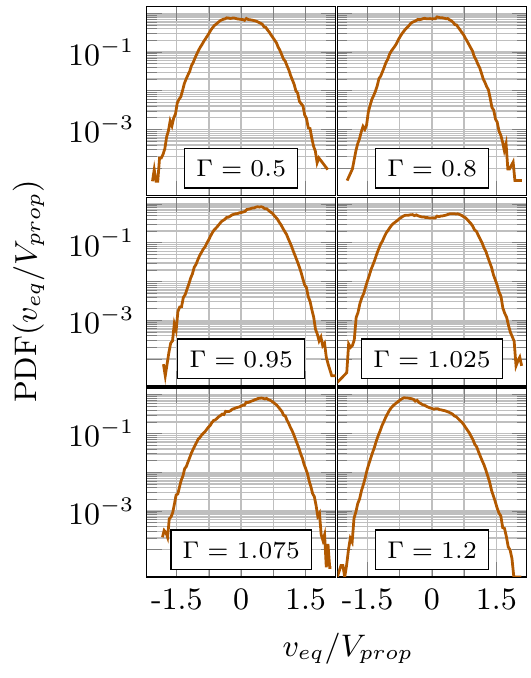}
\caption{Probability density functions of the normalized velocity for six different aspect ratios. They correspond to the time-series presented on the figure \ref{timeseries}.}
\label{PDFs}
\end{center}
\end{figure}

To confirm these observations we plot on the figure \ref{PDFs} the probability density functions (PDFs) corresponding to the previous time-series. Whereas for $\Gamma=0.5$ and $\Gamma=0.8$ the PDFs are symmetric, the ones corresponding to $\Gamma=0.95$ and $\Gamma=1.2$ are highly asymmetric. The bistability observed for $\Gamma=1.025$ corresponds to the double hump observed on the corresponding PDF. Finally whereas a bistability seemed to remain for $\Gamma=1.075$ the corresponding PDF reveals an asymmetric distribution. The bistability is not pronounced enough to have a signature on the PDF. The PDFs confirm the presence of a short window where the bistability can be observed and of two different behaviours before and after $\Gamma\approx 0.95$. A new question appears now: does exist a bifurcation which could explain the transition from symmetric to asymmetric PDFs when increasing $\Gamma$?

\begin{figure}[h!]
\begin{center}
\includegraphics[width=8cm]{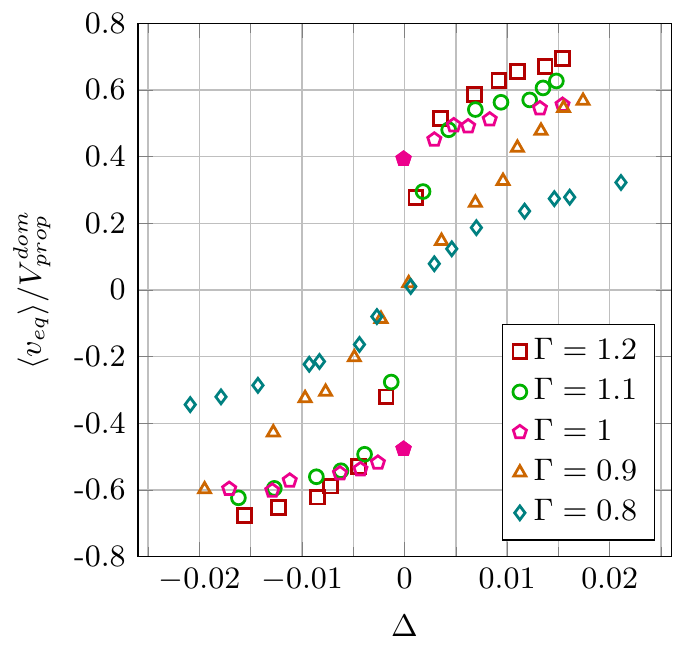}
\caption{Evolution of the averaged equatorial velocity with $\Delta$. $V_{prop}^{dom}$ represents the velocity of the dominant propeller. The filled symbols correspond to a bistability.}
\label{delta}
\end{center}
\end{figure}

To answer this question we look at the evolution of the equatorial velocity at the center of the cell while introducing a very slight difference of rotation frequency between the propellers. Figure \ref{delta} shows the equatorial velocity averaged over two minutes versus the asymmetry parameter $\Delta$ for five different aspect ratio. For $\Gamma=0.8$ and $\Gamma=0.9$ the evolution of $\langle v_{eq}\rangle$ is quite smooth. But for $\Gamma\geq 1$ we observe a very quick transition from a negative to a positive value of the equatorial velocity when $\Delta$ crosses 0. For a difference of rotation frequency of about 0.5\% ($\Delta=0.0025$) we observe a dramatic change of the sign of the averaged equatorial velocity. The system is the place of a bifurcation when increasing $\Gamma$: it transits from a smooth variation of $\langle v_{eq}\rangle$ with $\Delta$ to a very sudden change of the equatorial velocity when increasing $\Delta$. This explains why the PDFs of the velocity (figure \ref{PDFs}) are not symmetric for $\Gamma=0.95$ and $\Gamma=1.2$: the velocity imposed by the motors cannot be chosen with a sufficient precision to have exactly $\Delta=0$. Consequently for $\Gamma \geq0.95$, even if $\Delta$ is very close to 0, the shear layer is not in the middle of the cell. A new question appears: why do we observe such a bifurcation? A similar observation was made in a Von-K\'arm\'an swirling flow: the normalized and space-averaged angular momentum reveals the same behaviour with $\Delta$ by increasing the Reynolds number (from 150 to $8\times10^5$) \cite{cortet2010}.

\begin{figure}[h!]
\begin{center}
\includegraphics[width=8cm]{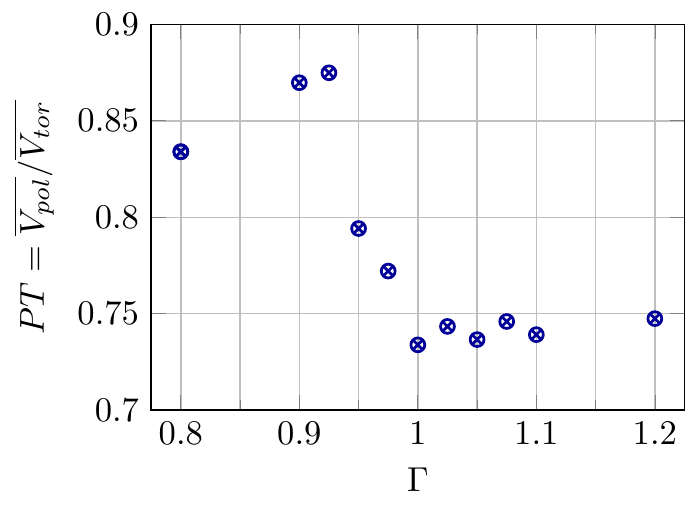}
\caption{Poloidal-toroidal ratio in the flow versus the aspect ratio for $\Delta\cong 0$ and $Re=2.9\times10^5$.}
\label{PT}
\end{center}
\end{figure}

We first investigate the structure of the mean flow. A good way consists in measuring the poloidal-toroidal ratio. We define it as $PT=\overline{V_{pol}}/\overline{V_{tor}}$ where $\overline{V_{pol}}$ and $\overline{V_{tor}}$ has been defined at eq. \ref{eq:V}. To estimate this ratio we perform a 2D mapping of the axial and equatorial velocities in the plane defined in the setup presentation (the axi-symmetry of the flow allows to this reduction of the measurement domain). We measure these two velocities over a mesh of about 2\,cm. The signals are averaged over 2\,minutes. Using the incompressibility of the flow we are able to obtain the third velocity component then to compute $PT$. The figure \ref{PT} shows the poloidal-toroidal ratio for different aspect ratios. For $\Gamma$ smaller than 0.8 the determination of $\overline{V_{pol}}$ and $\overline{V_{tor}}$ is not accurate for optical reasons (only a small part of the whole volume could be accessed) so we look at only the larger aspect ratios. We observe that for $\Gamma <0.95$, $PT$ remains higher than 0.8 whereas for larger aspect ratios it reaches a plateau close to 0.75. This is consistent with the bifurcation observed in the evolution of the shear layer position with $\Delta$. The mean flow structure could be at the origin of the bifurcation. However it cannot explain the appearance of the spontaneous long-time reversals: the poloidal-toroidal ratio is similar for $\Gamma=1$ (reversals) and $\Gamma=1.2$ (no reversal). We would like to signal that despite this behaviour, the mean flow is qualitatively similar to the one presented in figure 2 in ref. \cite{delatorre2007}. A careful analysis shows only a small deviation of the stagnation point for the poloidal component, and the maxima of the toroidal part, as well as the intensity of the poloidal part. 

These results suggest that the dynamics described above should be dependent on the propellers, as the shape of the blades controls the poloidal to toroidal ratio and their efficiency as kinetic energy injectors. We have performed some preliminary runs with two other kind of propellers: with eight similar curved blades and ten straight blades. We have verified that some of the characteristics presented here are still present, but for different values of the aspect ratio $\Gamma$.
\section{Langevin-based model} 

To improve the interpretation of these two phenomena (bifurcation and reversals) we propose a model for the shear layer position based on a Langevin equation. It is a very similar model to the one proposed in a different context by Machicoane \emph{et al.} \cite{machicoane2016}. We consider an unidimensional system where the shear layer can move along the $z$ direction and where a dimensionless potential $V$ can be defined in the cell.  The walls of the potential well represent the confinement induced by the propellers in the axial direction. The shear layer can move in this potential driven by random turbulent fluctuations. Its dynamics can be written as:

\begin{equation}
\frac{dz}{dt}=-\frac{dV}{dz} + B_0s(z)\eta(t).
\label{langevin}
\end{equation}

\noindent The last term represents the turbulent fluctuations where $B_0$ is the fluctuations intensity, $\eta(t)$ a coloured noise and $s(z)$ a spatial distribution of the turbulent fluctuations. We consider only the inertial turbulent scales range to influence the long-time dynamics of the shear layer. The coloured noise $\eta(t)$ is modeled as an Ornstein-Uhlenbeck process:

\begin{equation}
\frac{d\eta}{dt} = -\frac{\eta(t)}{\tau_p} + \sqrt{\frac{2}{\tau_p}}\xi(t),
\label{noise}
\end{equation}

\noindent where $\tau_p$ is a correlation time and $\xi(t)$ is a Gaussian noise such as $\langle\xi(t)\xi(t')\rangle=\delta(t-t')$.

We have seen that the bifurcation of the evolution of the shear layer position with $\Delta$ seems to be linked to the mean flow structure (figure \ref{PT}), but the spontaneous reversals between the two bifurcated symmetrical solutions have a different origin. Both features are included in this model.

\begin{figure}
\begin{center}
\includegraphics[width=8cm]{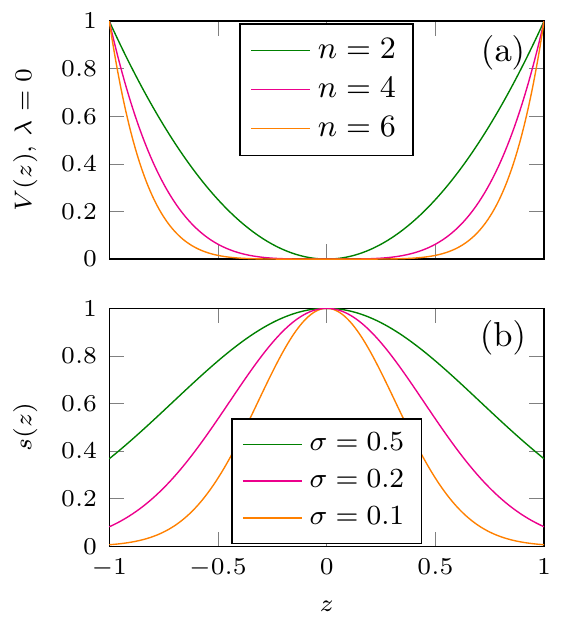}
\caption{Scheme of (a) the fluctuations spatial distributions for different $\sigma$ and of (b) the potential shape for different $n$ and $\lambda=0$. The $z$ coordinate and $V(z)$ are normalized here.}
\label{distribution}
\end{center}
\end{figure}

We propose to use a wide potential without energy barrier that mimics the confinement imposed by the propellers that forces the shear layer to remain in the central part of the cell (it corresponds to the second potential shape used by Machicoane \emph{et al.} \cite{machicoane2016}). We have $V(z)=|z|^n/n - \lambda z$ where $n\geq2$ can be turned continuously and $\lambda$ is used to break the symmetry along the $z$-axis (similarly  to the experimental parameter $\Delta$). Figure \ref{distribution}\,(a) shows the potential shape changes when we increase $n$. When the distance between the propellers increases, it is consistent to observe a flatter and flatter central region, stiffer and stiffer walls in the potential $V$, so to increase $n$ is similar to rise $\Gamma$.

On the other side, a naive view of the setup will suggest that the shear stress in the cell would be proportional to the velocity of the propellers' rim and inversely proportional to the distance between them: $V_{prop} / H \sim f_{prop} D/H  \sim  f_{prop} / \Gamma$. This would be correct if the shear were uniformly distributed along the $z$-axis. Nevertheless, the averaged velocity fields obtained in this configuration show that there is a layer (the shear layer) where this stress is mostly concentrated in a narrow region in the collision region of both recirculation cells. 
The turbulent fluctuations are more important in the shear layer than in the other zones of the flow. Consequently we propose a Gaussian spatial distribution of the turbulent fluctuations centered on $z=0$: $s(z)=\exp(-z^2/2\sigma^2)$ where $\sigma$ is the typical spatial extension of the turbulent fluctuations. Figure \ref{distribution}\,(b) shows the shape of the turbulent fluctuations Gaussian distribution. When $\sigma$ decreases the spatial extension of the fluctuations decreases too. The shear layer thickness is assumed to be constant because it is determined by the contra-rotating flow cells velocity which does not depend on the aspect ratio. Consequently the relative size of the shear layer decreases when the distance between the propellers (and so $\Gamma$) is increased. In our model decreasing $\sigma$ corresponds to increase the aspect ratio.

\begin{figure}
\begin{center}
\includegraphics[width=8cm]{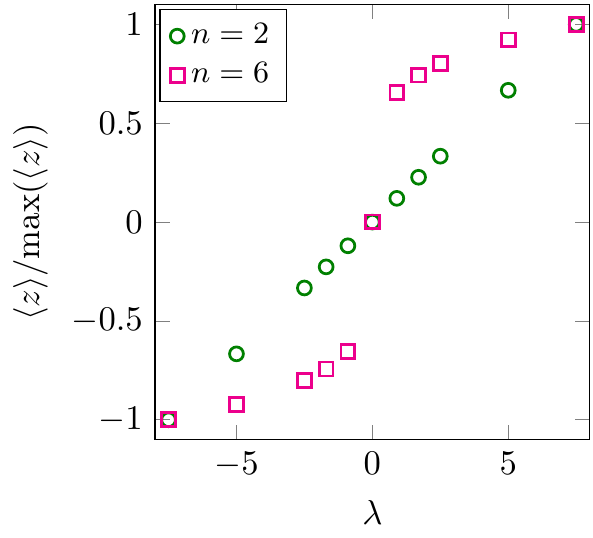}
\caption{Evolution of the shear layer position with $\lambda$ for different $n$ and $\sigma=0.1$. The normalization is made using the maximal $z$ in the explored range of $\lambda$.}
\label{transition_simu}
\end{center}
\end{figure}

We perform simulations of the shear layer position using this model. We fix $B_0$ and $\tau_p$ to 1. These parameters do not qualitatively change the following results. We present on the figure \ref{transition_simu} the evolution of the shear layer position with $\lambda$ for two different $n$ and $\sigma=0.1$. For $n=2$ the evolution of the shear layer is smooth with $\lambda$ whereas for $n=6$ (so for a larger aspect ratio) the transition is very sharp. This is very similar to the experimental behaviour and consistent with the assertion that this bifurcation is directly linked to the mean flow structure.

By watching the PDFs of the shear layer position for $\lambda$ close to 0 we are able to determine if the flow is bistable or not. We compute a phase diagram $(\sigma,n)$ to understand the behaviour of our system when changing $\Gamma$. We observe that for $n$ up to about 2.5 the behaviour of the shear layer position is smooth with $\lambda$. The bifurcation to a sharp evolution appears close to $n=2.5$. The zone where spontaneous reversals occur corresponds to high $\sigma$ where the spatial extension of the turbulent fluctuations is large relatively to the propellers span $H$. We propose two trajectories in the phase diagram which could be compared to what we observe in the experiment when we increase the aspect ratio. We can summarize the system history when we increase $\Gamma$ using the phase diagram: (i) the shear layer position is quite linear with $\Delta$, the bistability could be the one observed for $\Gamma=0.5$; (ii) the flow becomes bistable with a symmetry breaking when $\Delta$ is slightly changed; (iii) the bistability disappears but the sharp evolution of the shear layer position with $\Delta$ persists. This scenario is in quite good agreement with the experimental observations. Nevertheless two aspects seem to differ from our model: for $\Gamma=0.8$ we do not observe bistability, and we have a symmetry breaking without clear reversals for $\Gamma=0.95$. However a whole description of the experimental results is out of the scope of this model. Our objective was to identify the mechanisms responsible for the two main features: the bifurcation of the shear layer position and the spontaneous jumps between the solutions.

\begin{figure}
\begin{center}
\includegraphics[width=8cm]{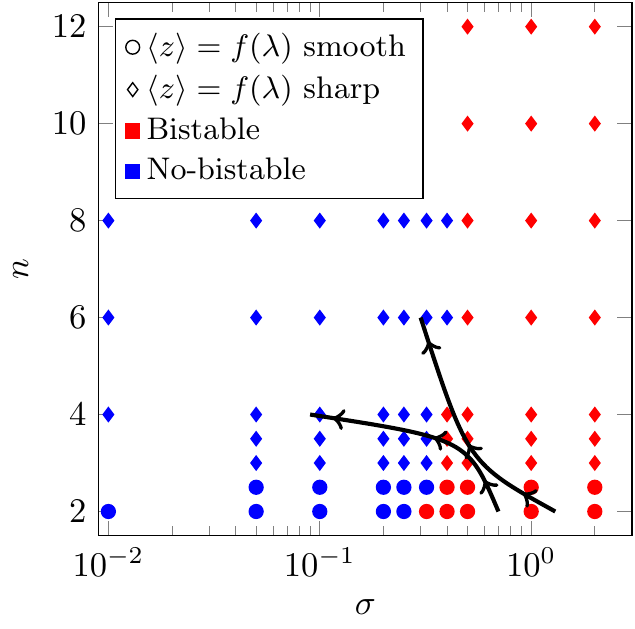}
\caption{Phase diagram of the shear layer behaviour based on the Langevin equation with coloured noise equation for $B_0=1$ and $\tau_p=1$. The black arrows are possible trajectories in the phase diagram when $\Gamma$ is increased.}
\label{diagramme}
\end{center}
\end{figure}

\section{Discussion and conclusion} 

To our knowledge this is the first systematic study about the influence of the aspect ratio on a Von-K\'arm\'an swirling flow. Although symmetry breaking with long-time reversals restoring the symmetry has been observed in this cell for $\Gamma=1$ \cite{delatorre2007}, we have shown that this phenomenon is highly dependent on the aspect ratio. For small aspect ratios ($\Gamma\lesssim 0.95$) the shear layer position reveals a smooth evolution with $\Delta$. Then a bifurcation appears and this evolution becomes very sharp. This bifurcation seems to be linked to a change of the mean flow structure as the measurement of the poloidal-toroidal ratio has shown. Simultaneously we have observed spontaneous long-time reversals in a strait window around $\Gamma=1$. Moreover, some short-time reversals have been revealed for $\Gamma$ around 0.5. 

To understand such phenomena we have proposed a model based on a Langevin equation inspired by a previous study \cite{machicoane2016}. The shear layer has been considered to move in a potential where the stiffness of the walls
% whose shape
varies with the aspect ratio. The turbulent fluctuations have been modeled with a coloured noise modulated by a Gaussian spatial distribution. We have succeeded in qualitatively reproduce the experimental observations. This provides a physical interpretation of the phenomena highlighted in this paper. When the distance between the propellers rises, the potential where the shear layer moves has a larger and larger flat zone. This allows the symmetry breaking and the stabilization of the shear layer at a position different from the center of the cell. Moreover, we have proposed the assumption that the shear layer thickness (and so on the typical absolute spatial extension of the turbulent fluctuations) does not depend on the distance between the propellers but only on the two fluid cells rotation velocity. Consequently when the aspect ratio is increased the relative spatial extension of the turbulent fluctuations decreases. A large relative spatial extension of the turbulent fluctuations could facilitate the passage of the shear layer from a side of the cell center to the other. 

Nevertheless, some disparities between the experimental observations and the Langevin-based model have appeared. Beyond the lack of resolution of the phase diagram (figure \ref{diagramme}), it probably exists a coupling between the mean flow structure and the turbulent fluctuations spatial extension. We have seen that the poloidal-toroidal ratio is not monotonic for $\Gamma<0.9$ (figure \ref{PT}). The consequence could be that the shear layer absolute thickness is not exactly constant and so that the evolution of $\sigma$ is not monotonic with $\Gamma$. It could explain why we do not observe bistability around $\Gamma=0.8$. Finally the coupling between the potential shape and the turbulent fluctuations could be an explanation of the difference between the reversals observed around $\Gamma=0.5$ (typical time of some tens of seconds) and the one around $\Gamma=1$ (typical time of one thousand seconds).

A consequence of this work is that the dynamics observed in the previous paper by de la Torre {\em et al.} \cite{delatorre2007} can be reproduced without the introduction of any potential barrier that separates both symmetry breaking states. In that paper there was no explanation for the origin of the proposed barrier, and using the model of this study, whose characteristics are closely related to the experimental properties, its necessity disappears. This model should recover the behaviour of a large set of problems where turbulent flows are restricted to confined geometries. One example can be the destabilization of a turbulent wake behind an axisymmetric obstacle that breaks some symmetries of the problem \cite{rigas2015}, and whose dynamics was described using a model close to the one presented in \cite{delatorre2007}.

%One example can be the destabilization of a turbulent wake that breaks some symmetries of the problem \cite{rigas2015}, and whose dynamics was described using a model close to the one presented in ref. \cite{delatorre2007}.

\bibliography{biblio_ArXiv_Liot}

\end{document}